\documentclass[letterpaper, 10pt, conference]{ieeeconf}

\IEEEoverridecommandlockouts  
\usepackage{amsmath, amssymb, amsfonts}
\usepackage{dsfont}
\usepackage{graphicx}
\usepackage{float}
\usepackage{color}
\usepackage[noadjust]{cite}

\usepackage{graphicx} 
\usepackage{quiver}
\usepackage{verbatim}
\usepackage{todonotes}
\usepackage{xcolor}

\usepackage{amsmath, amsfonts, amssymb, amsthm, mathtools}
\usepackage{stmaryrd}
\let\labelindent\relax
\usepackage{enumitem}
\usepackage{thmtools}
\usepackage{fancyhdr}
\usepackage{hyperref}
\usepackage{tikz}
\usepackage{comment}
\usepackage[normalem]{ulem}
\usepackage{booktabs}
\usepackage{subcaption}  
\usepackage[font=footnotesize]{caption}  
\usepackage{mathrsfs}
\usepackage{bbm}
\usepackage{adjustbox}

\newcommand{\id}{\mathrm{id}}
\newcommand{\define}[1]{\emph{#1}}

\newtheorem{theorem}{Theorem}

\newtheorem{lemma}{Lemma}
\newtheorem{proposition}{Proposition}
\theoremstyle{definition}
\newtheorem{assumption}{Assumption}
\newtheorem{definition}{Definition}

\newtheorem{example}{Example}
\newtheorem{remark}{Remark}

\newcommand{\cat}[1]{\mathsf{#1}}
\newcommand{\fun}[1]{\mathcal{#1}}
\newcommand{\Vect}{\cat{Vect}}
\newcommand{\CellSh}{\cat{CellSh}}
\newcommand{\narr}[1]{\mathscr{#1}}
\usepackage{quiver}

\title{\LARGE \bf A Temporal Cellular Sheaf Framework for Multi-Agent Systems with Switching Communication Topologies
}

\author{
Brandon Fallin$^{1}$,
Wilmer Leal$^{1}$,
Austin Copeland$^{1}$,
Benjamin Merlin Bumpus$^{2}$, \\
James Fairbanks$^{1}$,
Warren E. Dixon$^{3}$
\thanks{$^{1}$Brandon Fallin, Wilmer Leal, Austin Copeland, and James Fairbanks are with the Department of Mechanical and Aerospace Engineering at the University of Florida, Gainesville, FL, USA. Emails: {\texttt{ \{brandonfallin, wleal, acopeland1, fairbanksj\}@ufl.edu}}. BF, WL, AC, and WD were supported by AFRL (FA8651-24-1-0018), AFOSR (FA9950-19-1-0169), and ARL (W911NF-25-2-0045). JF was supported by DARPA (HR0011222038) and ONR (N00014-23-1-2339). Any opinions, findings, and conclusions or recommendations are those of the authors and do not necessarily reflect the views of the sponsoring agency.} 
\thanks{$^{2}$Benjamin Merlin Bumpus is with the Institute of Mathematics and Statistics at the University of S\~ao Paulo, S\~ao Paulo, SP, Brazil. Email: {\texttt {bumpus@usp.br}}. Benjamin Merlin Bumpus was supported by the S\~ao Paulo Research Foundation (FAPESP), grant 2025/16921-5.}
\thanks{$^{3}$Warren E. Dixon is with the College of Engineering at Virginia Tech, Blacksburg, VA, USA. Email: {\texttt{wdixon@vt.edu}}.}
}

\begin{document}
\maketitle
\thispagestyle{empty}
\pagestyle{empty}

\begin{abstract}
	Cellular sheaves serve as a natural modeling framework for encoding heterogeneity in the state and interaction spaces of multi-agent systems. This work examines switching communication graphs with the use of temporal cellular sheaves, which we construct to represent heterogeneous agent state spaces, the maps that relate them, and their respective communication constraints. We then develop sufficient conditions under which a multi-agent system defined on a temporal cellular sheaf can solve the target tracking problem by generalizing the existing concept of joint connectivity. Using a Lyapunov-based stability analysis, we demonstrate that the target tracking error of the network converges to a neighborhood of the origin.
\end{abstract}

\section{Introduction}
A central objective in the study of multi-agent systems (MASs) is the design of decentralized or distributed control laws which ensure coordination and system stability, provided the communication topology meets certain constraints~\cite{Yu.Xia2012,Wen.Duan.ea2013,Xu.Cao.ea2014}. 
Heterogeneity is commonly represented in these systems through nonidentical agent dynamics, accounting for agents operating in different domains. Agents that operate in different state spaces or contain different sensing modalities offer another form of heterogeneity. Networks of this form are commonly constructed via matrix-weighted approaches which equip graph interactions (edges) with matrix-valued weights, representing varied communication or sensing relationships~\cite{Trinh.VanNguyen.ea2018,Hansen2021,Pan.Shao.ea2022,Wang.Pan.ea2024,Zhang.Pan.ea2024}.

Standard graph-based models of MASs are limited in that they do not provide a framework which simultaneously captures heterogeneity in agent dynamics, local-to-global consistency constraints, and compositional structure. Moreover, extending such models to time-varying settings introduces additional difficulties, since both the communication topology and coordination constraints may evolve over time. Cellular sheaves provide a formalism for addressing these challenges~\cite{Hansen.Ghrist2019}. 

Using a cellular sheaf framework, a communication network can be represented by a graph, and vector spaces can be assigned to both nodes (agents) and edges (interactions). Linear maps encode sensing, communication, or compatibility constraints between agents, allowing users to model both individual and joint state spaces in a unified structure. Matrix-weighted graphs and bearing-based approaches can be recovered as a particular subclass of these representations (e.g., \cite{Trinh.VanNguyen.ea2018,Hansen2021,Zhao.Zelazo2016,VanTran.Park.ea2018}). Cellular sheaves have emerged as a natural modeling framework across a diverse range of applications, including the diffusion of information in social networks~\cite{Hansen.Ghrist2021, Bosca.Ghrist2026a}, neural network inference and training~\cite{Bosca.Ghrist2026b}, multi-agent coordination~\cite{Hanks.Riess.ea2025}, leader-following consensus~\cite{Hanks.Nino.ea2026}, the modeling of hybrid dynamical systems \cite{Moustafa.Abdelghany2026}, and asynchronous distributed optimization~\cite{Zhao.Hanks.ea2026}. Cellular sheaves generalize the classical graph Laplacian to accommodate for heterogeneous agent state spaces and linear edge-based restriction maps~\cite{Hanks.Nino.ea2026, Hanks.Riess.ea2025}. In particular, cellular sheaves connect multi-agent coordination problems with tools from computational algebraic topology~\cite{Edelsbrunner.Harer2010}. 

Sheaf Laplacians and cohomology offer mechanisms to analyze information propagation and global coordination in distributed systems. For instance,~\cite{Hanks.Nino.ea2026} formulates heterogeneous multi-agent coordination over a fixed communication network as a harmonic extension problem, where target states serve as boundary data. The resulting decentralized control law yields exponential convergence to a neighborhood of the unique sheaf-consistent global state. Despite these advances, existing cellular sheaf-based approaches to multi-agent coordination primarily assume a fixed communication topology~\cite{Hansen.Ghrist2021,Hanks.Nino.ea2026}. In practice, the communication topology of a MAS may vary over time due to sensing constraints, environmental interference, or hardware failure. To this end, we develop a framework for cellular sheaf MASs with switching communication topologies. Previous work has explored formation control and leader-following under switching communication topologies for matrix-weighted graphs~\cite{Wang.Pan.ea2024} and bearing-based approaches~\cite{Tang.Loria2022,Cheng.Yan.ea2026}, but to the best of our knowledge, this problem  remains open for cellular sheaves. In this work, we make the following contributions:

\begin{itemize}
    \item \textit{Temporal cellular sheaves for switching communication systems.} We leverage existing notions of a category of cellular sheaves (e.g., \cite{Curry2014,Ghrist.Riess2020,Hansen2020,Felber.Flores.ea2025}) which we denote by $\CellSh$. Using this construction, we model multi-agent systems with switching communication topologies as temporal cellular sheaves (see Fig.~\ref{fig:temporal-cellsheaf}), defined as $\CellSh$-valued cumulative narratives~\cite{Bumpus.Leal.ea2025}.
    \item \textit{Stability guarantees.} We perform a Lyapunov stability analysis which leverages the methodology from~\cite{Meng.Chen.ea2018} to establish conditions under which each agent's target tracking error converges to a neighborhood of the origin.
\end{itemize} 

\section{Notation and Preliminaries}
Denote the set of agents by $V_q\triangleq \{1,2,\ldots,N \}$ and the target agent by $\tau$. Define the vertex set $V\triangleq V_q\cup \{\tau\}$. Let $E_{qq}\subseteq \{\{i,j\} \mid i,j\in V_q \}$ denote the set of agent-agent edges and let $E_{q\tau}\subseteq \{ \{i,\tau\} \mid i\in V_q \}$ denote the set of agent-target edges. The set $E\triangleq E_{qq}\cup E_{q\tau}$ describes every edge that may be realized on a subsequently defined cellular sheaf. Denote the admissible communication graph by $G\triangleq(V,E)$ and the neighborhood of agent $i$ by $\mathcal{N}_i\triangleq\{ j\in V\mid\{i,j\}\in E \}$ for all $i\in V$. 
For a set $S$ and an element $s$, let $\mathbf{1}_S(s)$ denote the indicator function. The agent-agent adjacency matrix $A\triangleq[a_{ij}]\in\mathbb{R}^{N\times N}$ has entries $a_{ij}= \mathbf{1}_{E_{qq}}(\{i,j\})$, and the agent-target indicator $b_i$ is given by $b_i\triangleq \mathbf{1}_{E_{q\tau}}(\{i,\tau\})$. For any subgraph $G^\prime = (V^\prime, E^\prime)\subseteq G$, let $V(G^\prime)\triangleq V^\prime$ denote the node set of $G^\prime$ and $E(G^\prime)\triangleq E^\prime$ denote the edge set of $G^\prime$. Let $\mathds{1}_n$ denote the vector of $1$s in $\mathbb{R}^n$.
For a collection of vector spaces $\{V_1,\ldots,V_m\}$, denote their direct sum by $\oplus_{i\in\{1,\ldots,m\}}V_i$. For a set $A$, let $\operatorname{int}(A)$ denote its interior. A map $f:A\to B$ is of class $\mathscr{C}^k(A,B)$ if it is $k$-times continuously differentiable.

\section{Cellular Sheaves for Multi-Agent Systems}\label{sec:cellular-sheaves}
\subsection{Communication Topologies as Categories}
Modeling communication topologies as categories allows us to encode inter-agent connectivity, define consistent state assignments, and define coherent maps between MASs.

\begin{definition}[Category]\label{def:category}
    A category $\cat{C}$ consists of objects, morphisms $f:X\to Y$ between objects, an identity morphism $\id_X:X\to X$ for each object, and an associative composition law $g\circ f$  whenever the codomain of $f$ equals the domain of $g$, such that $f\circ \id_X=f$ and $\id_Y\circ f=f$.
\end{definition}

\begin{example}[Category of vector spaces]\label{ex:vect}
    The category $\Vect_k$ has vector spaces over a fixed field $k$ as objects and linear maps as morphisms. Identities are identity maps, and composition is the usual composition of linear maps.
\end{example}

\begin{example}[Time category~\cite{Bumpus.Leal.ea2025}]\label{ex:time-category}
    The category $\cat{I}$ (resp. $\cat{I}_{\mathbb{N}}$) has closed intervals $[\ell,\ell']\subseteq \mathbb{R}$ (resp. ${\mathbb{N}}$) as objects, with a unique morphism $[\ell,\ell']\to [a,b]$ whenever $[\ell,\ell']\subseteq [a,b]$. Composition is given by inclusion. A \define{time category} \(\cat{T}\) is any sub-join-semilattice of either \(\cat{I}\) or \(\cat{I}_\mathbb{N}\).
\end{example}

\begin{definition}\label{def:incidence-graph}
    (Incidence category of a graph).
    Let $G=(V,E)$ be a finite, undirected graph. Its \emph{incidence category}, denoted $\mathrm{Inc}(G)$, is the category whose objects are the cells of $G$ (nodes and edges) and whose non-identity morphisms are the unique incidence maps $i\to e$ whenever $i$ is an endpoint of $e$; no other nontrivial compositions exist. Fig.~\ref{fig:cellular-sheaf-components} depicts a graph $G$ and its incidence category $\mathrm{Inc}(G)$.
\end{definition}

We model the communication topology of a multi-agent system by the incidence category $\mathrm{Inc}(G)$ of its underlying communication graph $G$. In this representation, objects correspond to agents and their communication links, while morphisms encode the incidence relations between them.

\subsection{States and Interaction Spaces as Functors}
\begin{figure}[hbtp]
    \centering
    \includegraphics[width=\columnwidth]{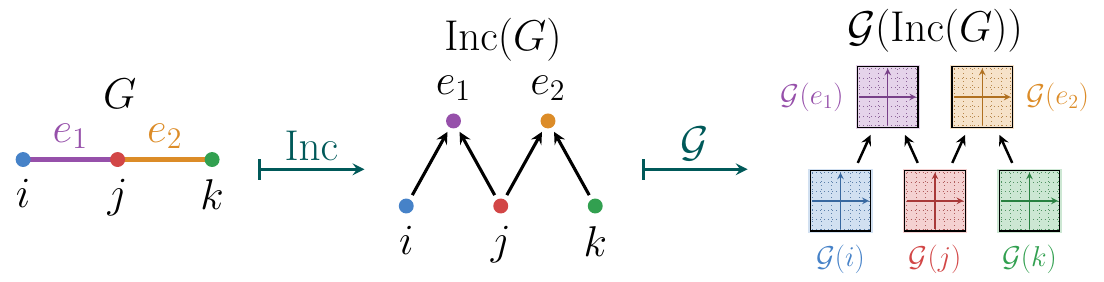}
    \caption{The communication network $G$ is represented by a graph whose nodes represent agents and whose edges represent communication links. Its incidence category $\mathrm{Inc}(G)$ encodes nodes and edges as objects and incidences as morphisms. A cellular sheaf $\fun{G}:\mathrm{Inc}(G)\to\Vect_k$ assigns state spaces to agent and interaction objects, together with linear maps associated with incidence morphisms, yielding the diagram $\fun{G}(\mathrm{Inc}(G))$. 
    }
    \label{fig:cellular-sheaf-components}
\end{figure}
We now formalize the second map in Fig.~\ref{fig:cellular-sheaf-components}: each agent $i\in \mathrm{Inc}(G)$ is assigned a state space $\fun{G}(i)$ which we refer to as a~\emph{node stalk}, each edge $e$ a joint interaction space $\fun{G}(e)$ which we refer to as an~\emph{edge stalk}, and each incidence $i\to e$ a linear map $\fun{G}(i\to e)$ which we refer to as a~\emph{restriction map}. To assign these data consistently over the communication topology $\mathrm{Inc}(G)$, we use a functor.

\begin{definition}[Covariant functor]
    Let $\cat{C}$ and $\cat{D}$ be categories. A covariant functor $\fun{F}:\cat{C}\to\cat{D}$ assigns to each object $X$ of $\cat{C}$ an object $\fun{F}(X)$ of $\cat{D}$ and to each morphism $f:X\to Y$ in $\cat{C}$ a morphism $\fun{F}(f):\fun{F}(X)\to \fun{F}(Y)$ in $\cat{D}$ such that $\fun{F}(\id_X)=\id_{\fun{F}(X)}$ and $\fun{F}(g\circ f)=\fun{F}(g)\circ \fun{F}(f)$ whenever $g\circ f$ is defined.
\end{definition}

Functors specify how objects and morphisms are mapped while preserving the behavior encoded by composition. The class of functors of interest in this work is introduced next.

\begin{definition}[Cellular sheaf]
\label{def:cellular-sheaf-graph}
    A \emph{cellular sheaf on a graph $G$} is a covariant functor $\fun{G} : \mathrm{Inc} (G) \longrightarrow \Vect_k$.
\end{definition}

Definition~\ref{def:cellular-sheaf-graph} establishes \emph{vector-valued cellular sheaves}, which we refer to simply as \emph{cellular sheaves} throughout this work. A cellular sheaf $\fun{G}:\mathrm{Inc} (G)\to\Vect_k$ specifies admissible states and compatibility constraints imposed by the communication topology $\mathrm{Inc} (G)$.

\subsection{The Category of Cellular Sheaves}
While a cellular sheaf $\fun{F}:\mathrm{Inc} (F)\to\Vect_k$ encodes a single MAS, our goal is to model how this system transforms into another $\fun{G}$ $:\mathrm{Inc} (G)\to\Vect_k$ as the communication topology or the associated state and interaction spaces change. To formalize this transformation, we define morphisms between cellular sheaves that simultaneously relate the communication topologies and the assigned state data.  We first consider MASs $\fun{G}_1,\fun{G}_2:\mathrm{Inc} (G)\to\Vect_k$ sharing the same communication topology. Comparing these systems amounts to relating their agent and interaction vector spaces via a collection of coherent linear maps, called a \emph{natural transformation}.

\begin{definition}[Natural transformation for cellular sheaves]
    Let $\fun{G}_1,\fun{G}_2:\mathrm{Inc} (G)\to\Vect_k$ be cellular sheaves on the same graph $G$. A \emph{natural transformation} $\alpha:\fun{G}_1\Rightarrow\fun{G}_2$ consists of  linear maps $\alpha_c:\fun{G}_1(c)\to\fun{G}_2(c)$, one for each object $c$ (agent or edge) of $\mathrm{Inc}(G)$, such that for every incidence morphism $f:i\to e$, we have $\alpha_e \circ \fun{G}_1(f) = \fun{G}_2(f) \circ \alpha_i$.
\end{definition}

The components $\alpha_i:\fun{G}_1(i)\to\fun{G}_2(i)$ of the natural transformation relate the state spaces assigned to agent $i$ by the two MASs, while $\alpha_e:\fun{G}_1(e)\to\fun{G}_2(e)$ relates their interaction spaces. The naturality condition $\alpha_e\circ\fun{G}_1(f)=\fun{G}_2(f)\circ\alpha_i$ ensures that these transformations are compatible with the communication (restriction) maps of the two systems. Specifically, transforming an agent's state from $\fun{G}_1$ to $\fun{G}_2$ via $\alpha_i$ and then applying the communication map $\fun{G}_2(f)$ yields the same result as first applying $\fun{G}_1(f)$ within $\fun{G}_1$ and then transforming the resulting interaction data to $\fun{G}_2$ via $\alpha_e$. 

Next, we use a functor $\fun{T}:\mathrm{Inc} (F)\to\mathrm{Inc} (G)$ which allows us to compare two MASs $\fun{F}$ and $\fun{G}$ with different communication topologies.

\begin{definition}[Morphisms of cellular sheaves]
\label{def:cellular-sheaf-morphism}
    Let $\fun{F}:\mathrm{Inc}(F)\to\Vect_k$ and $\fun{G}:\mathrm{Inc}(G)\to\Vect_k$ be cellular sheaves. A \emph{morphism of cellular sheaves} $(\fun{T},\alpha):\fun{F}\to\fun{G}$ consists of a functor $\fun{T}:\mathrm{Inc}(F)\to\mathrm{Inc}(G)$ relating their communication topologies and a natural transformation $\alpha:\fun{F}\Rightarrow\fun{G}\circ\fun{T}$ relating their assigned vector spaces, as illustrated in the diagram
    \[
    \begin{tikzcd}
        {\mathrm{Inc}(F)} && {\mathrm{Inc}(G)} \\
        & {\Vect_k}
        \arrow["{\fun{T}}", from=1-1, to=1-3]
        \arrow[""{name=0, anchor=center, inner sep=0}, "{\fun{F}}"', from=1-1, to=2-2]
        \arrow[""{name=1, anchor=center, inner sep=0}, "{\fun{G}}", from=1-3, to=2-2]
        \arrow["\alpha", Rightarrow, nfold, from=0, to=1]
    \end{tikzcd}
    \mathpunct{.}
    \]
\end{definition}
The functor $\fun{T}$ determines how agents and interactions in $\fun{F}$ correspond to those in $\fun{G}$. The natural transformation $\alpha$ then specifies coherent maps between state and interaction spaces. Next, we construct the category of cellular sheaves which we denote by $\CellSh$. Prior notions of a category of cellular sheaves include: (i) a functor category from a fixed poset category (typically the cells of a complex ordered by inclusion) to $\Vect_k$~\cite{Curry2014,Ghrist.Riess2020}, (ii) a category whose objects are cellular sheaves on cell complexes and whose morphisms are commuting squares~\cite{Felber.Flores.ea2025}, and (iii) a category whose objects are pairs consisting of an undirected graph and an associated cellular sheaf, and whose morphisms are pairs consisting of a graph morphism and a compatible sheaf morphism~\cite{Hansen2020}.

\begin{proposition}[Category of cellular sheaves]\label{prop:category-cellular-sheaves}
    The category $\CellSh$ has as objects cellular sheaves $\fun{G}:\mathrm{Inc} (G)\to \Vect_k$ on finite graphs $G$, and as morphisms those of Definition~\ref{def:cellular-sheaf-morphism}.
\end{proposition}
\begin{proof}
    Composition is defined componentwise. Given morphisms $(\fun{T}_1,\alpha):\fun{F}\to\fun{G}$ and $(\fun{T}_2,\beta):\fun{G}\to\fun{H}$, their composite is $(\fun{T}_2,\beta)\circ(\fun{T}_1,\alpha)\triangleq(\fun{T}_2\circ\fun{T}_1,\lambda_c)$, where for each cell $c$ of the graph $F$ in \(\fun{F}:\mathrm{Inc} (F)\to \Vect_k\), the component of the natural transformation is given by $\lambda_c\triangleq\beta_{(\fun{G}\circ\fun{T}_1)(c)} \circ \alpha_c : \fun{F}(c) \to (\fun{H}\circ\fun{T}_2\circ\fun{T}_1)(c)$. The identity morphism on $\fun{F}$ is $(\id_{\mathrm{Inc} (F)},\id_{\fun{F}})$. Associativity and the identity laws follow from the associativity of functor composition and composition of linear maps.
\end{proof}

\section{Temporal Cellular Sheaves}\label{sec:temporal-cellsheaves}
\begin{figure}[hbtp]
    \centering
    \includegraphics[width=\columnwidth]{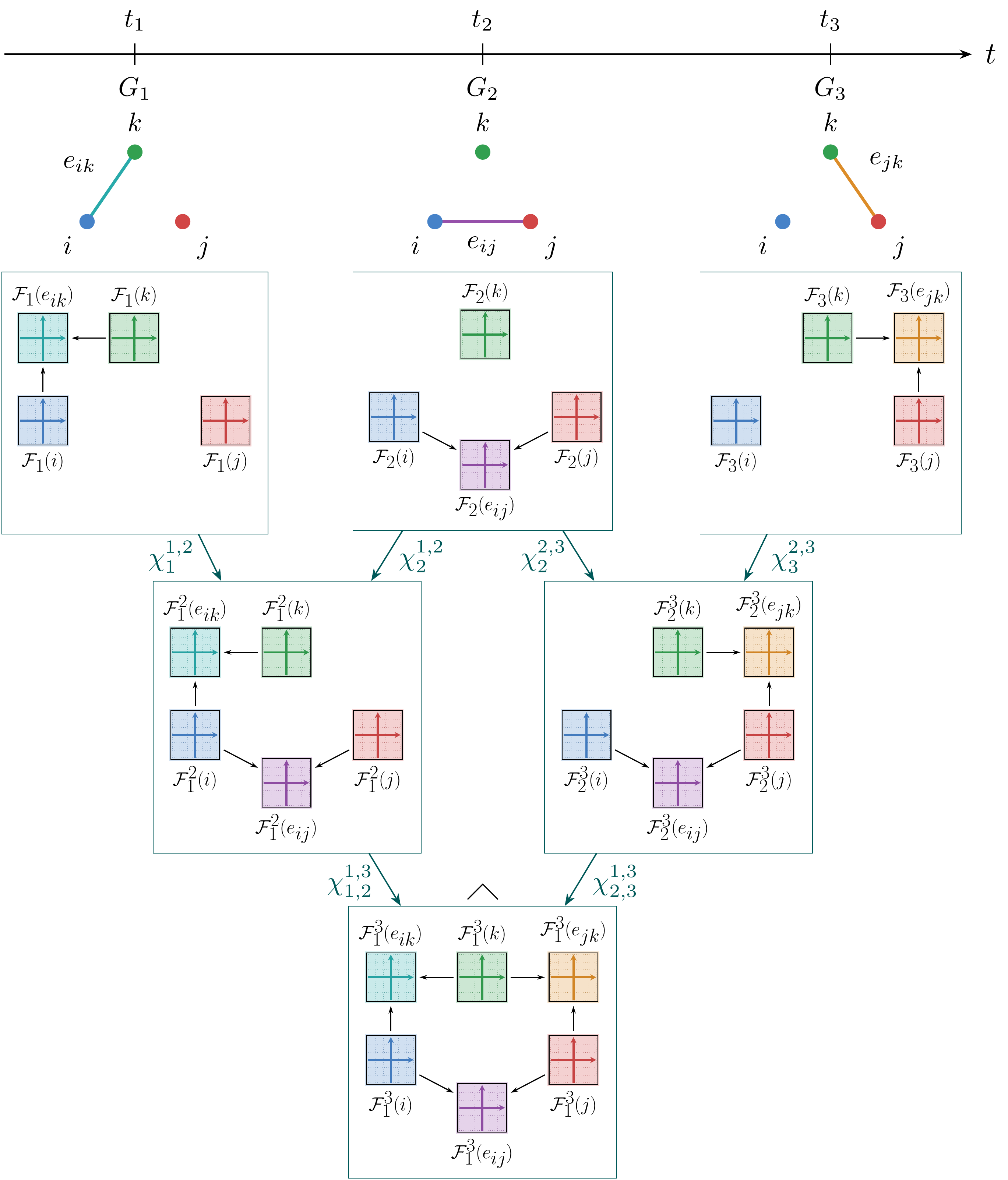}
    \caption{Temporal cellular sheaf model for a multi-agent system with a switching communication topology. From top to bottom, the figure shows: (i) the time axis with steps $t_1,t_2,t_3$, (ii) the graphs $G_1,G_2,G_3$, illustrating the time-varying communication topology, (iii) the cellular sheaves $\mathcal{F}_1,\mathcal{F}_2,\mathcal{F}_3$ defined on the incidence categories of $G_1,G_2,G_3$, (iv) morphisms $\chi_1^{1,2},\chi_2^{1,2},\chi_2^{2,3},\chi_3^{2,3}$ between successive cellular sheaves encoding how the system structure evolves across communication topologies, and (v) a cellular sheaf $\mathcal{F}_1^3$ obtained as a pushout (gluing) of $\mathcal{F}_1^2$ and $\mathcal{F}_2^3$.}
    \label{fig:temporal-cellsheaf}
\end{figure}
Our goal is to model multi-agent systems whose communication topologies vary over time while retaining the ability to reason about the cumulative communication structure and the agent states observed over larger time intervals. This requires a mechanism for gluing MASs defined on smaller intervals into a single system defined on their union. Categorically, such gluing is captured by pushouts.

\begin{definition}[Pushout]\label{def:pushout}
	Let $\cat{C}$ be a category and consider morphisms $A \xleftarrow{f} C \xrightarrow{g} B$ in $\cat{C}$. 
	A \emph{pushout} of this diagram consists of an object $P$ together with morphisms $A\xrightarrow{i_1} P \xleftarrow{i_2} B$ such that $i_1\circ f= i_2 \circ g$. Moreover, $P$ satisfies the following universal property: for any object $X$ with morphisms $j_1:A\to X$ and $j_2:B\to X$ such that $j_1\circ f = j_2\circ g$, there exists a unique morphism $h:P\to X$ such that $h\circ i_1 = j_1$ and $h\circ i_2 = j_2$.
\end{definition}

The category $\CellSh$ provides a general framework for relating cellular sheaf MASs. For the temporal constructions considered here, however, we need to model the accumulation of MASs over time intervals. This motivates restricting the morphisms of $\mathrm{IncGrph}$, which denotes the category of incidence categories, to those induced by graph inclusions, leading to the following category.

\begin{definition}[Category of incidence categories]\label{def:category-of-incidence}
    The category $\mathrm{IncGrph}_{\mathrm{emb}}$ has as objects the incidence categories of finite, undirected graphs and as morphisms the functors $\mathrm{Inc}(F)\hookrightarrow\mathrm{Inc}(G)$ induced by graph inclusions $F\hookrightarrow G$, meaning that the vertex and edge sets of $F$ are subsets of the corresponding vertex and edge sets of $G$ and all incidence relations are preserved. Since these morphisms are inclusions, there is at most one morphism between any two objects. Thus, $\mathrm{IncGrph}_{\mathrm{emb}}$ is a \emph{thin category}.
\end{definition}

$\mathrm{IncGrph}_{\mathrm{emb}}$ admits a principled notion of gluing.

\begin{lemma}\label{lem:incgrph-pushouts}
    The category $\mathrm{IncGrph}_{\mathrm{emb}}$ admits pushouts.
\end{lemma}
\begin{proof}
    $\mathrm{IncGrph}_{\mathrm{emb}}$ is thin. Given a span $\mathrm{Inc}(F)\xhookleftarrow{}\mathrm{Inc}(H)\xhookrightarrow{}\mathrm{Inc}(G),$ the union graph $F\cup G$ is the least graph containing both $F$ and $G$. Therefore, $\mathrm{Inc}(F\cup G)$ is the least upper bound of the span and hence its pushout.
\end{proof}

The pushout construction above lifts from communication topologies to cellular sheaves. We therefore define the subcategory $\CellSh_{\mathrm{emb}}$ whose objects are cellular sheaves and whose morphisms are cellular sheaf morphisms $(\fun{T},\alpha)$ for which $\fun{T}$ is induced by a graph inclusion. In $\CellSh_{\mathrm{emb}}$, pushouts provide a categorical notion of union for MASs.

\begin{lemma}\label{lem:cellsh-emb-pushout}
	The category $\CellSh_{\mathrm{emb}}$ admits pushouts.
\end{lemma}
\begin{proof}
    Let $\fun F \xleftarrow{(\fun T_1,\alpha_1)} \fun H \xrightarrow{(\fun T_2,\alpha_2)} \fun G$ be a span in $\CellSh_{\mathrm{emb}}$. At a high level, the proof proceeds by first computing a pushout of the underlying communication topologies, yielding the incidence category on which the pushout cellular sheaf will be defined. We then use left Kan extensions to transport the cellular sheaves in the span to this common incidence category and construct the pushout in the corresponding functor category. 
    
    By Lemma~\ref{lem:incgrph-pushouts}, the underlying incidence categories admit a pushout $\mathrm{Inc}(F\cup G)$. Since $\Vect_k$ is cocomplete and the incidence categories are small, the left Kan extensions $\mathrm{Lan}_{\fun L}\fun F$, $\mathrm{Lan}_{\fun R}\fun G$, and $\mathrm{Lan}_{\fun K}\fun H$ exist, where $\fun K=\fun L\circ\fun T_1=\fun R\circ\fun T_2$. The morphisms $\alpha_1$ and $\alpha_2$ induce a span $\mathrm{Lan}_{\fun L}\fun F \xleftarrow{\bar\alpha_1} \mathrm{Lan}_{\fun K}\fun H \xrightarrow{\bar\alpha_2} \mathrm{Lan}_{\fun R}\fun G$ in $[\mathrm{Inc}(F\cup G),\Vect_k]$, which admits a pushout $\fun P$ since functor categories having a cocomplete codomain are cocomplete~\cite[\S 8.5.1]{Schubert1972}. The universal properties of the pushout incidence category, the left Kan extensions, and $\fun P$ yield the required pushout in $\CellSh_{\mathrm{emb}}$. Therefore, $\CellSh_{\mathrm{emb}}$ admits pushouts.
\end{proof}

We now instantiate the temporal framework of \cite{Bumpus.Leal.ea2025,Leal.Bumpus.ea2026} to cellular sheaves. We use the category $\cat{I}_{\mathbb{N}}$ from Example~\ref{ex:time-category} as our model of discrete time and assign data to intervals using copresheaves.

\begin{definition}[Copresheaf]
    Let $\cat{C}$ and $\cat{D}$ be categories. A \emph{copresheaf} on $\cat{C}$ with values in $\cat{D}$ is a covariant functor $\hat{F}:\cat{C}\to\cat{D}$. Thus, every morphism $U\rightarrow V$ in $\cat{C}$ induces a morphism $\hat{F}(U)\to\hat{F}(V)$ in $\cat{D}$.
\end{definition}

To describe how a MAS evolves over time, we need to specify how larger time intervals can be decomposed into smaller ones from which the system can be reconstructed. Covering families provide this notion of temporal decomposition. In particular, we use the Johnstone coverage~\cite{Johnstone1999}, which covers an interval by two subintervals meeting at a common time point.

\begin{definition}[Johnstone coverage]
The \emph{Johnstone coverage} on $\cat{I}_{\mathbb{N}}$ assigns to each interval $[\ell,\ell']$ the covering family \( J([\ell,\ell'])=\{([\ell,p],[p,\ell']) \mid \ell\le p\le \ell'\}. \)
\end{definition}

The category of time intervals together with this notion of covering families, $(\cat{I}_{\mathbb{N}},J)$, is called a \emph{site}. We seek a temporal model in which the system assigned to a longer interval is determined by the systems assigned to adjacent subintervals. This model is formalized by cumulative narratives~\cite{Bumpus.Leal.ea2025}. 

\begin{definition}[Cumulative narrative~{\cite{Bumpus.Leal.ea2025}}]\label{def:t-sheaf}
    Let $\cat{T}$ be a time category equipped with the Johnstone coverage, and let $\cat{D}$ be a category with pushouts. A \emph{$\cat{D}$-valued cumulative narrative} on $\cat{T}$ is a copresheaf $\hat{F}:\cat{T}\to\cat{D}$ such that for every interval $[\ell,\ell']$ and every cover $([\ell,p],[p,\ell'])$, $\hat{F}([\ell,\ell']) = \hat{F}([\ell,p]) +_{\hat{F}([p,p])} \hat{F}([p,\ell'])$.
\end{definition}

Since Lemma~\ref{lem:cellsh-emb-pushout} shows that $\CellSh_{\mathrm{emb}}$ has pushouts, Definition~\ref{def:t-sheaf} applies with $\cat{D}=\CellSh_{\mathrm{emb}}$. This yields the main notion of this section.

\begin{definition}[Temporal cellular sheaf]\label{def:temp-cellsheaf}
    Let $\cat{T}$ be a time category equipped with the Johnstone coverage. A \emph{temporal cellular sheaf} is a $\CellSh_{\mathrm{emb}}$-valued cumulative narrative $\narr{T}:\cat{T}\to\CellSh_{\mathrm{emb}}$.
\end{definition}

In the context of multi-agent systems, the pushout condition in Definition~\ref{def:temp-cellsheaf} says that the system over the interval $[\ell,\ell']$ is obtained by taking the union of the systems active on the subintervals $[\ell,p]$ and $[p,\ell']$, provided they agree at the shared switching time $p$. The temporal cellular sheaf structure formalizes the MAS with switching communication topology depicted in Fig.~\ref{fig:temporal-cellsheaf} and will later be used to define union cellular sheaves and formulate a sheaf-theoretic generalization of joint connectivity.

\section{Problem Formulation}\label{sec:problem-formulation}
Given a subgraph $G^\prime\subseteq G$, denote the space of $0$-cochains of $G^\prime$ by $C^0 (G^\prime;\mathcal{F})\triangleq \oplus_{i\in V(G^\prime)}\mathcal{F}(i)$, and denote the space of $1$-cochains of $G^\prime$ by $C^1 (G^\prime;\mathcal{F})\triangleq \oplus_{\{i,j\}\in E(G^\prime)}\mathcal{F}(\{i,j\})$. Equip each undirected edge $\{i,j\}\in E$ with the arbitrary orientation $i\to j$ and define the coboundary operator $\delta^0_{\mathcal{F}}:C^0 (G;\mathcal{F})\to C^1 (G;\mathcal{F})$, which measures disagreement on edges. For $x\in C^0 (G;\mathcal{F})$, the coboundary operator acts edgewise as $(\delta_{\mathcal{F}}^{0} x)_{ij}=\mathcal{F}_{i\trianglelefteqslant ij}(x_{i})-\mathcal{F}_{j\trianglelefteqslant ij}(x_{j})$, where $\mathcal{F}_{i\trianglelefteqslant ij}:\mathcal{F}(i)\to\mathcal{F}(\{i,j\})$ denotes the restriction map from node $i$ to edge $\{i,j\}$. The sheaf Laplacian $L_{\mathcal{F}}:C^0 (G;\mathcal{F})\to C^0 (G;\mathcal{F})$ is defined as $L_{\mathcal{F}}\triangleq \left( \delta_{\mathcal{F}}^{0} \right)^\top \delta_{\mathcal{F}}^{0}$~\cite{Hansen.Ghrist2021}.

\subsection{Extension of Target States}\label{subsec:extension}
In consensus and coordination problems, the objective is to yield states $x\in C^0(G;\mathcal{F})$ which satisfy $L_{\mathcal{F}}x=0$. To formalize this objective, partition the coboundary operator into a component from the agent nodes and a component from the target as $\delta_{\mathcal{F}}^0=[ \delta_q,\delta_\tau ]$, where $G_q\triangleq (V_q, E_{qq})\subseteq G$ denotes the agent-agent subgraph, $\delta_q:C^0(G_q;\mathcal{F})\to C^1(G;\mathcal{F})$, and $\delta_\tau:\mathcal{F}(\tau)\to C^1(G;\mathcal{F})$. Applying this decomposition to the sheaf Laplacian yields the \emph{target extension} problem given by
\begin{align}\label{eq:target-extension}
	\begin{bmatrix}
		\delta_q^\top \delta_q & \delta_q^\top \delta_\tau \\
		\delta_\tau^\top \delta_q & \delta_\tau^\top \delta_\tau
	\end{bmatrix}
	\begin{bmatrix}
		x^\ast \\
		x_{\tau}
	\end{bmatrix} =
	\begin{bmatrix}
		0 \\ 
		0
	\end{bmatrix},
\end{align}
where we seek a unique $x^\ast\in C^0(G_q;\mathcal{F})$ satisfying~\eqref{eq:target-extension} for each $x_{\tau}\in \mathcal{F}(\tau)$, where $[(x^{\ast})^\top,x_\tau^\top]^\top$ is called a \emph{global section} of $\mathcal{F}$. Next, we provide conditions under which~\eqref{eq:target-extension} may be solved, yielding a similar result to the conditions derived in~\cite[Proposition A.1]{Bosca.Ghrist2026a}. 
\begin{lemma}\label{lem:image-condition}
	Suppose $\delta_q^\top\delta_q \succ 0$ and fix $x_{\tau}\in \mathcal{F}(\tau)$. Then, there exists a unique $x^\ast\in C^0(G_q;\mathcal{F})$ satisfying~\eqref{eq:target-extension} if and only if $\delta_\tau x_{\tau}\in\mathrm{Im}\ \delta_q$. 
\end{lemma}
\begin{proof}
    Let $x^\ast\in C^0(G_q;\mathcal{F})$ satisfy~\eqref{eq:target-extension}. Then,
    \begin{align}\label{eq:harmonic-extension}
        x^\ast=-\left( \delta_q^\top\delta_q \right)^{-1}\delta_q^\top \delta_\tau x_{\tau}.
    \end{align}
    Since $\delta_q^\top\delta_q \succ 0$, its inverse exists. Further, the relation
    \begin{align}\label{eq:second-relation}
        \delta_\tau^\top \delta_q x^\ast+\delta_\tau^\top\delta_\tau x_{\tau} = 0
    \end{align}
    holds. Define $P_q\triangleq \delta_q\left( \delta_q^\top\delta_q \right)^{-1}\delta_q^\top$. Substituting~\eqref{eq:harmonic-extension} into~\eqref{eq:second-relation} yields $\delta_\tau^\top\left( I - P_q \right)\delta_\tau x_{\tau}=0$, where $P_q:C^1(G;\mathcal{F})\to C^1(G;\mathcal{F})$ is an orthogonal projection onto $\mathrm{Im}\ \delta_q$. Left multiplying by $x_{\tau}^\top$ yields $x_{\tau}^\top\delta_\tau^\top (I-P_q)\delta_\tau x_{\tau}=0$, implying that $\delta_\tau x_{\tau}$ and $(I-P_q)\delta_\tau x_\tau$ are orthogonal. Since $I-P_q$ is an orthogonal projection onto $(\mathrm{Im}\ \delta_q)_\perp$, $\delta_\tau x_{\tau}\in\mathrm{Im}\ \delta_q$. 
    
    Suppose $\delta_\tau x_{\tau}\in\mathrm{Im}\ \delta_q$. Since $\delta_q$ is of full column rank, there exists a unique $x\in C^0(G_q;\mathcal{F})$ satisfying $\delta_\tau x_{\tau}=\delta_q x$. Letting $x=-x^\ast$ and left multiplying by $\delta_q^\top$ yields $\delta_q^\top\delta_q x^\ast + \delta_q^\top\delta_\tau x_{\tau}=0$. Then, since $I-P_q$ is an orthogonal projection onto $(\mathrm{Im}\ \delta_q)_\perp$, $(I-P_q)\delta_\tau x_{\tau}=0$. Left multiplying by $\delta_\tau^\top$ yields $\delta_\tau^\top (I-P_q)\delta_\tau x_{\tau}=0$. Performing algebraic manipulation, substituting the definition of $P_q$, and substituting the relation $\delta_q^\top\delta_\tau x_{\tau}=-\delta_q^\top\delta_q x^\ast$ yields $\delta_\tau^\top\delta_q x^\ast + \delta_\tau^\top\delta_\tau x_{\tau}=0$. Thus, \eqref{eq:target-extension} holds. 
    
    For uniqueness, suppose $x_1^\ast,x_2^\ast$ satisfy~\eqref{eq:target-extension}. Then, $\delta_q^\top\delta_q (x_1^\ast - x_2^\ast)=0$. Since $\delta_q$ is of full column rank, $\mathrm{ker}\ \delta_q^\top\delta_q=\{ 0 \}$ and $x_1^\ast=x_2^\ast$. Therefore, $x^\ast$ is unique.
\end{proof}

\begin{remark}\label{rem:constant-sheaf-extend}
    In the standard target tracking framework (e.g., \cite{Hong.Hu.ea2006}, \cite{Ni.Cheng2010}), the hypotheses of Lemma~\ref{lem:image-condition} are satisfied under mild conditions. Specifically, consider the case in which $G_q\subseteq G$ is connected and there exists $i\in V(G_q)$ such that $\{i,\tau\}\in E$. Let $\mathcal{H}:\mathrm{Inc}(G)\to\Vect_k$ denote the constant cellular sheaf on $\mathbb{R}^n$ where $\mathcal{H}(v)=\mathbb{R}^n$ for all $v\in V$, $\mathcal{H}(\{i,j\})=\mathbb{R}^n$ for all $\{i,j\}\in E$, and $\mathcal{H}_{i\trianglelefteqslant ij}=I_n$ for every incident node-edge pair. Then, for each $x_{\tau}\in\mathbb{R}^n$, it can be shown that $x^\ast\in\oplus_{i\in V(G_q)}\mathbb{R}^n$ exists, is unique, and $(x^\ast)_i = x_{\tau}$ for all $i\in V_q$. 
\end{remark}

\subsection{Switching Cellular Sheaves}\label{subsec:switching}
To quantify the resulting temporal cellular sheaf, we generalize the concept of joint connectivity in~\cite{Meng.Chen.ea2018}. Let $\mathscr{S}\triangleq\{1,2,\ldots, S\}$ denote the finite set of switching modes where $S\in\mathbb{Z}_{>0}$ and let $\sigma:\mathbb{R}_{\geq 0}\to\{ 1,2,\ldots,S \}$ denote a piecewise constant switching signal, indicating which mode is active at a given time. 

Let each mode correspond to a subgraph of $G$ satisfying $G_s\subseteq G$ and $V(G_s)=V(G)$ for all $s\in\mathscr{S}$. The cellular sheaf associated with each subgraph is $\mathcal{F}_s=\mathcal{F}\vert_{G_s}$, where $\mathcal{F}\vert_{G_s}$ denotes the cellular sheaf $\mathcal{F}$ restricted to the subgraph $G_s$. We say that $\mathcal{F}_s$ is a \emph{sub-cellular sheaf} of $\mathcal{F}$ and denote this relation by $\mathcal{F}_s\subseteq \mathcal{F}$. A sub-cellular sheaf of $\mathcal{F}$ is any cellular sheaf $\mathcal{F}_s$ having a morphism in $\CellSh_{\mathrm{emb}}$ $(\fun{T}_s,\alpha_s):\mathcal{F}_s\to\mathcal{F}$ for which $\alpha_s$ is the identity for all node and edge stalks in $\mathcal{F}_s$. Then, $\mathcal{F}_1,\ldots,\mathcal{F}_S\subseteq\mathcal{F}$. 

Consider the infinite sequence of increasing time instants given by $\{t_k\}_{k=0}^\infty$. Then, partition each of the intervals $[t_k,t_{k+1})$ using another finite sequence of increasing time instants given by $\{t_k^{\ell}\}_{\ell =0}^{s_k}$, where $(s_k -1)\in\mathbb{Z}_{\geq 0}$ denotes the number of switches which occur on the interval $[t_k,t_{k+1})$. Note that $t_k^{s_k}=t_{k+1}$. The subintervals $[t_{k}^{i},t_{k}^{j}),\ s_k\geq j>i\geq 0$ denote when a sub-cellular sheaf $\mathcal{F}_{\sigma (t)}$ is active. We make the following assumption regarding the sequence of sub-cellular sheaves.
\begin{assumption}[{\cite[Assumption 2.1]{Meng.Chen.ea2018}}]\label{ass:infinite-sequence}
	There exist constants $B_\sigma,\beta_\sigma\in\mathbb{R}_{>0}$ satisfying $t_{k+1}-t_k < B_\sigma$ for all $k\in\mathbb{Z}_{\geq 0}$ and $t_k^j-t_k^i > \beta_\sigma$ for all $s_k\geq j>i\geq 0$ and all $k\in\mathbb{Z}_{\geq 0}$.
\end{assumption}
Next, we characterize the structure of the sheaf Laplacian for a sub-cellular sheaf.
\begin{lemma}\label{lem:sub-cellular-sheaf}
	Let $G_s\subseteq G$ satisfy $V(G_s)=V(G)$ and denote its associated sub-cellular sheaf by $\mathcal{F}_s=\mathcal{F}\vert_{G_s}$. Then,
	\begin{align}\label{eq:sub-laplacian}
		L_{\mathcal{F}_s}=\begin{bmatrix}
			\delta_q^\top \Psi_s \delta_q & \delta_q^\top \Psi_s \delta_\tau \\ 
			\delta_\tau^\top \Psi_s \delta_q & \delta_\tau^\top \Psi_s \delta_\tau 
		\end{bmatrix},
	\end{align}
	where 
    $\Psi_s\triangleq\mathrm{blkdiag}\{ ( \mathbf{1}_{E(G_s)}(e) I_{\mathcal{F}(e)})_{e\in E(G)} \}$ for any fixed ordering of $E(G)$.
\end{lemma}
\begin{proof}
    There exists a map $\psi_{\mathcal{F}_s}:C^1(G;\mathcal{F})\to C^1(G_s;\mathcal{F}_s)$ defined for any $\xi\in C^1(G;\mathcal{F})$ as $\left( \psi_{\mathcal{F}_s} (\xi) \right)_e = (\xi)_e$ for all $e\in E(G_s)$. The map $\psi_{\mathcal{F}_s}$ omits the values of $\xi$ corresponding to edges $e\in E(G)\backslash E(G_s)$. The coboundary operator of $\mathcal{F}_s$ denoted by $\delta_{\mathcal{F}_s}^0:C^0(G_s;\mathcal{F}_s)\to C^1(G_s;\mathcal{F}_s)$ satisfies $\delta_{\mathcal{F}_s^0}=\psi_{\mathcal{F}_s}\delta_{\mathcal{F}}^0$. 
    
    Partition $\delta_{\mathcal{F}}^0=[\delta_q,\delta_\tau]$ and define $\Psi_s=\psi_{\mathcal{F}_s}^\top \psi_{\mathcal{F}_s}$. Applying the facts that $L_{\mathcal{F}_s}=\left( \delta_{\mathcal{F}_s}^0 \right)^\top \delta_{\mathcal{F}_s}^0$ and $\delta_{\mathcal{F}_s}^0=\psi_{\mathcal{F}_s}\delta_{\mathcal{F}}^0$ yields~\eqref{eq:sub-laplacian}. We see that for any $\xi_1,\xi_2\in C^1(G;\mathcal{F})$,
    \begin{align}\label{eq:identity-inner-prod}
        \langle \psi_{\mathcal{F}_s}\xi_1,\psi_{\mathcal{F}_s}\xi_2 \rangle = \sum_{e\in E(G_s)}\left(\xi_1\right)_e^\top I_{\mathcal{F}(e)} \left(\xi_2\right)_e.
    \end{align}
    For any fixed ordering of $E(G)$, the right-hand side of~\eqref{eq:identity-inner-prod} can be written as $\xi_1^\top \mathrm{blkdiag}\{( \mathbf{1}_{E(G_s)}(e) I_{\mathcal{F}(e)})_{e\in E(G)} \} \xi_2$. Applying the definition of $\Psi_s$ and the fact that $\xi_1,\xi_2$ are arbitrary yields $\Psi_s=\mathrm{blkdiag}\{ ( \mathbf{1}_{E(G_s)}(e) I_{\mathcal{F}(e)} )_{e\in E(G)} \}$. 
\end{proof}

Next, we define the union cellular sheaf.
\begin{definition}[Union cellular sheaf]
	Given a fixed, undirected graph $G=(V,E)$, a cellular sheaf $\mathcal{F}:\mathrm{Inc}(G)\to\Vect_k$, and a collection of sub-cellular sheaves $\{ \mathcal{F}_1,\ldots,\mathcal{F}_S \}$, we define the \emph{union cellular sheaf} as $\bigcup_{s\in\{1,\ldots ,S\}} \mathcal{F}_s \triangleq \mathcal{F}\vert_{\cup_{s\in\{1,\ldots ,S\}} G_s}$.
\end{definition}
We now use the union cellular sheaf to formulate a sheaf-theoretic generalization of joint connectivity.
\begin{definition}[Joint extensibility]
	Let the union cellular sheaf over an interval $[t_k,t_{k+1})$ be denoted by $\mathcal{F}_{\Sigma_k}\triangleq \bigcup_{t\in[t_k,t_{k+1})} \mathcal{F}_{\sigma(t)}$ having the union graph $G_{\Sigma_k}=\cup_{t\in[t_k,t_{k+1})}G_{\sigma (t)}$. Let the coboundary operator $\delta_{\mathcal{F}_{\Sigma_k}}^0 : C^0 (G;\mathcal{F})\to C^1 (G_{\Sigma_k};\mathcal{F}_{\Sigma_k})$ of $\mathcal{F}_{\Sigma_k}$ be partitioned as $\delta_{\mathcal{F}_{\Sigma_k}}^0 = [\delta_{q_{\Sigma_k}},\delta_{\tau_{\Sigma_k}}]$. An infinite sequence of sub-cellular sheaves is \emph{jointly extensible} if $\delta_{q_{\Sigma_k}}^\top \delta_{q_{\Sigma_k}} \succ 0$ for all $k\in\mathbb{Z}_{\geq 0}$.
\end{definition}
\begin{remark}
	In the standard target tracking setting on the constant cellular sheaf $\mathcal{H}$, joint connectivity implies joint extensibility. We note that joint connectivity for target tracking can be formulated by requiring both the union agent-agent graph $G_{q_{\Sigma_k}}$ and the union graph $G_{\Sigma_k}$ to be connected for every $k\in\mathbb{Z}_{\geq 0}$. If both $G_{q_{\Sigma_k}}$ and $G_{\Sigma_k}$ are connected, there exists at least one node in $G_{q_{\Sigma_k}}$ which is connected to the target agent. Therefore, $L_{G_{q_{\Sigma_k}}}+\mathrm{diag}\{ b_{1,\Sigma_k},\ldots,b_{N,\Sigma_k} \}\succ 0$. Applying Remark~\ref{rem:constant-sheaf-extend} yields $\delta_{q_{\Sigma_k}}^\top \delta_{q_{\Sigma_k}}\succ 0$, and any MAS defined on the constant cellular sheaf $\mathcal{H}$ whose agent-agent communication graph is jointly connected is also jointly extensible.
\end{remark}

\noindent The temporal cellular sheaf framework provides a mechanism for gluing cellular sheaf-based MASs  and for defining joint extensibility.
In this section, we use the temporal cellular sheaf framework in a context where gluing assembles subsystems of a larger MAS. However, we note that the notion of gluing established by the pushout in $\CellSh_{\mathrm{emb}}$ is considerably more general.

\subsection{Error System}\label{subsec:error-system}
In this section, we establish the target tracking problem for a cellular sheaf MAS having a time-varying communication topology. We write $\mathbb{R}^{n_{i}}$ to denote $\mathcal{F}(i)$ for all $i\in V$ and $\mathbb{R}^{n_{ij}}$ to denote $\mathcal{F}(\{i,j\})$ for all $\{i,j\}\in E$, where $n_{i}\triangleq\mathrm{dim}\ \mathcal{F}(i),\ n_{ij}\triangleq\mathrm{dim}\ \mathcal{F}(\{i,j\})$. The dynamics of the target agent are
\begin{align}\label{eq:target-dynamics}
	\dot{x}_{\tau}=f_{\tau}(x_{\tau},t),
\end{align}
where $x_{\tau},\dot{x}_{\tau}\in\mathbb{R}^{n_{\tau}}$ denote the target agent's position and velocity, respectively, and $f_{\tau}:\mathbb{R}^{n_{\tau}}\times \mathbb{R}_{\geq 0}\to\mathbb{R}^{n_{\tau}}$ denotes the target agent's unknown drift dynamics. We make the following assumption with respect to the target agent.
\begin{assumption}\label{ass:target-bounds}
	There exist constants $\bar{x}_{\tau},\bar{\dot{x}}_{\tau}\in\mathbb{R}_{>0}$ satisfying $\lVert x_{\tau} (t) \rVert \leq \bar{x}_{\tau}$ and $\lVert \dot{x}_{\tau} (t) \rVert\leq \bar{\dot{x}}_{\tau}$ for all $t\in\mathbb{R}_{\geq 0}$.
\end{assumption}
The agent dynamics are
\begin{align}\label{eq:agent-dynamics}
	\dot{x}_i = f_i(x_i,t)+g_i(x_i,t)u_i + \omega_i (t),
\end{align}
where $x_i,\dot{x}_i\in\mathbb{R}^{n_i}$ denote agent $i$'s position and velocity, respectively, $f_i:\mathbb{R}^{n_i}\times\mathbb{R}_{\geq 0}\to\mathbb{R}^{n_i}$ denotes the unknown, heterogeneous drift dynamics, $g_i:\mathbb{R}^{n_i}\times\mathbb{R}_{\geq 0}\to\mathbb{R}^{n_i\times m_i}$ denotes a known control effectiveness matrix, $u_i\in\mathbb{R}^{m_i}$ denotes the control input, and $\omega_i:\mathbb{R}_{\geq 0}\to\mathbb{R}^{n_i}$ denotes an unknown, time-varying disturbance. We make the following assumptions to ensure the well-posedness of the target tracking problem.
\begin{assumption}\label{ass:drift}
	The drift dynamics $f_i$ are of class $\mathscr{C}^1(\mathbb{R}^{n_i}\times\mathbb{R}_{\geq0}, \mathbb{R}^{n_i})$ and the maps $t\mapsto f_i(x_i,t)$, $t\mapsto\nabla f_i(x_i,t)$ are uniformly bounded in $t$ when $x_i\in\Omega_i$ for all $i\in V(G)$, where $\Omega_i\subset\mathbb{R}^{n_i}$ is any compact subset of $\mathbb{R}^{n_i}$. 
\end{assumption}
\begin{assumption}\label{ass:control-eff}
	The control effectiveness matrix $g_i$ is of full row rank and is of class $\mathscr{C}^0 (\mathbb{R}^{n_i}\times\mathbb{R}_{\geq 0},\mathbb{R}^{n_i\times m_i})$ for all $i\in V(G_q)$. Further, for each $x_i\in\mathbb{R}^{n_i}$, the map $t\mapsto g_i(x_i,t)$ is uniformly bounded in $t\in\mathbb{R}_{\geq 0}$ for all $i\in V(G_q)$.  	
\end{assumption}
\begin{assumption}\label{ass:disturbance}
	The unknown disturbance $\omega_i$ is of class $\mathscr{C}^0(\mathbb{R}_{\geq 0},\mathbb{R}^{n_i})$ and satisfies $\lVert \omega_i (t) \rVert\leq \bar{\omega}$ for all $t\in \mathbb{R}_{\geq 0}$ and $i\in V(G_q)$, where $\bar{\omega}\in\mathbb{R}_{>0}$.
\end{assumption}
\begin{assumption}[{\cite[Assumption 6]{Hanks.Nino.ea2026}}]\label{ass:relative-cohomology}
	The relative cohomology of $G$ with respect to $\{ \tau \}$ is trivial. That is, $H^0 (G,\{ \tau \};\mathcal{F})=0$.
\end{assumption}
\begin{assumption}\label{ass:satisfies-image}
	The cellular sheaf $\mathcal{F}$ satisfies $\mathrm{Im}\ \delta_\tau\subseteq\mathrm{Im}\ \delta_q$.
\end{assumption}
\begin{assumption}\label{ass:jointly-extensible}
	The infinite sequence of sub-cellular sheaves specified by the switching signal $\sigma(t)$ is jointly extensible.
\end{assumption}

Define a target tracking error $e\in C^0(G_q;\mathcal{F})$ such that
\begin{align}\label{eq:ensemble-target-tracking-error}
	e\triangleq x^\ast - x,
\end{align}
where $x\in C^0(G_q;\mathcal{F})$ is defined as $x\triangleq\oplus_{i\in V(G_q)} x_i$. Each agent constructs an estimate of the target extension. Denote this estimate by $\hat{x}_i\in\mathbb{R}^{n_i}$ for all $i\in V(G_q)$ and define a target extension estimation error as 
\begin{align}\label{eq:ensemble-extension-error}
	\tilde{x}\triangleq x^\ast -\hat{x},
\end{align}
where $\hat{x}\in C^0(G_q;\mathcal{F})$ is defined as $\hat{x}\triangleq\oplus_{i\in V(G_q)}\hat{x}_i$. Then, each agent will attempt to regulate its state to its estimate of the target extension using its control input. This objective is encoded using a target extension regulation error as 
\begin{align}\label{eq:ensemble-regulation-error}
	\hat{e}\triangleq\hat{x}-x,
\end{align}
where $\hat{e}=\oplus_{i\in V(G_q)}\hat{e}_i$. 

\section{Control Design}\label{sec:control-design}
\subsection{Observer Update Law}\label{subsec:observer-update-law}
Based on the subsequent stability analysis, the update law for the target extension observer is designed as
\begin{align}\label{eq:observer}
	\begin{split}
		\dot{\hat{x}}_i &= k_1 \sum_{j\in V(G_q)} a_{ij}^{\sigma (t)} \mathcal{F}_{i\trianglelefteqslant ij}^\top \left( \mathcal{F}_{j\trianglelefteqslant ij} \hat{x}_j - \mathcal{F}_{i\trianglelefteqslant ij} \hat{x}_i \right) \\
		&+ k_1 b_i^{\sigma(t)} \mathcal{F}_{i\trianglelefteqslant i\tau}^\top \left( \mathcal{F}_{\tau\trianglelefteqslant i\tau} x_{\tau} - \mathcal{F}_{i\trianglelefteqslant i\tau} \hat{x}_i \right), 
	\end{split}
\end{align}
for all $i\in V(G_q)$, where $k_1\in\mathbb{R}_{>0}$ denotes a user-selected gain value, $a_{ij}^{\sigma(t)}:\mathbb{R}_{\geq 0}\to \{ 0,1 \}$ satisfies $a_{ij}^{\sigma(t)}=1$ if $\{i,j\}\in E(G_{\sigma(t)})$ for $i,j\in V(G_q)$ and $a_{ij}^{\sigma(t)}=0$ otherwise, and $b_i^{\sigma (t)}:\mathbb{R}_{\geq 0}\to \{ 0,1 \}$ satisfies $b_i^{\sigma (t)}=1$ if $\{i,\tau \}\in E(G_{\sigma(t)})$ for $i\in V(G_q)$ and $b_i^{\sigma (t)}=0$ otherwise. Given a sub-cellular sheaf $\mathcal{F}_{\sigma(t)}$, partition its coboundary operator as $\delta_{\mathcal{F}_{\sigma (t)}}=[\delta_{q_{\sigma (t)}},\delta_{\tau_{\sigma (t)}}]$. Then, substituting the ensemble form of~\eqref{eq:observer} into the time derivative of~\eqref{eq:ensemble-extension-error} and applying the definition of $x^\ast$ yields
\begin{align}\label{eq:x-tilde-dot-sub}
	\begin{split}
		\dot{\tilde{x}}&= -k_1\delta_{q_{\sigma (t)}}^\top \delta_{q_{\sigma (t)}} \tilde{x} - \big( \delta_q^\top\delta_q \big)^{-1}\delta_q^\top\delta_\tau \dot{x}_{\tau} \\
		&-k_1 \delta_{q_{\sigma (t)}}^\top \Big( \delta_{q_{\sigma (t)}} \big( \big( \delta_q^\top\delta_q \big)^{-1} \delta_q^\top \delta_\tau \big) - \delta_{\tau_{\sigma (t)}} \Big) x_{\tau}.
	\end{split}
\end{align}
Applying the facts that $\delta_{q_{\sigma (t)}}^\top \delta_{q_{\sigma (t)}}=\delta_q^\top \Psi_{\sigma (t)} \delta_q$ and $\delta_{q_{\sigma (t)}}^\top \delta_{\tau_{\sigma (t)}}=\delta_q^\top \Psi_{\sigma (t)} \delta_\tau$ to~\eqref{eq:x-tilde-dot-sub} yields
\begin{align}\label{eq:x-tilde-dot-extension}
	\begin{split}
		\dot{\tilde{x}}&= -k_1\delta_{q_{\sigma (t)}}^\top \delta_{q_{\sigma (t)}} \tilde{x} - \big( \delta_q^\top\delta_q \big)^{-1}\delta_q^\top\delta_\tau \dot{x}_{\tau} \\
		&-k_1 \delta_{q}^\top \Psi_{\sigma (t)} \left( P_q - I \right) \delta_\tau x_{\tau}.
	\end{split}
\end{align}
By Assumptions~\ref{ass:relative-cohomology} and~\ref{ass:satisfies-image}, Lemma~\ref{lem:image-condition} is satisfied for all $x_{\tau}\in\mathbb{R}^{n_{\tau}}$, yielding $(I-P_q)\delta_\tau x_{\tau} (t)=0$ for all $t\in\mathbb{R}_{\geq 0}$. Applying this fact to~\eqref{eq:x-tilde-dot-extension} gives
\begin{align}\label{eq:ensemble-estimation-error-dynamics}
	\dot{\tilde{x}}=-k_1\delta_{q_{\sigma (t)}}^\top\delta_{q_{\sigma (t)}}\tilde{x} - \left( \delta_q^\top\delta_q \right)^{-1}\delta_q^\top\delta_\tau \dot{x}_{\tau}. 
\end{align}

\subsection{Control Law}\label{subsec:control-law}
The control law is designed as
\begin{align}\label{eq:controller}
	u_i = g_i^{\dagger} (x_i,t) \big( k_2\hat{e}_i + \dot{\hat{x}}_i \big),
\end{align}
where $k_2\in\mathbb{R}_{>0}$ is a user-selected gain value and $g_i^{\dagger}$ denotes the Moore-Penrose pseudoinverse of $g_i$. We make the following assumption of $g_i^{\dagger}$ to ensure that the control input remains bounded and continuous. 
\begin{assumption}\label{ass:control-eff-inv}
    The Moore-Penrose pseudoinverse of the control effectiveness matrix $g_i^{\dagger}$ is of class $\mathscr{C}^0(\mathbb{R}^{n_i}\times\mathbb{R}_{\geq 0}, \mathbb{R}^{m_i\times n_i})$, and for each $x_i\in\mathbb{R}^{n_i}$, the map $t\mapsto g_i^{\dagger}(x_i,t)$ is uniformly bounded in $t\in\mathbb{R}_{\geq 0}$ for all $i\in V(G_q)$.
\end{assumption}
Define $\omega:\mathbb{R}_{\geq 0}\to C^0(G_q;\mathcal{F})$ as $\omega(t)\triangleq\oplus_{i\in V(G_q)}\omega_i (t)$ and $f:C^0(G_q;\mathcal{F})\times\mathbb{R}_{\geq 0}\to C^0 (G_q;\mathcal{F})$ as $f(\mu,t)\triangleq\oplus_{i\in V(G_q)}f_i (\mu_i,t)$ for any $\mu\in C^0(G_q;\mathcal{F})$. Taking the time derivative of~\eqref{eq:ensemble-regulation-error} and substituting the ensemble form of~\eqref{eq:controller} into the ensemble form of~\eqref{eq:agent-dynamics} gives
\begin{align}\label{eq:ensemble-regulation-error-dynamics}
	\dot{\hat{e}}=-k_2\hat{e} - \left( f(x,t) -f(\hat{x},t) \right)-\omega(t)-f(\hat{x},t).
\end{align}

\section{Stability Analysis}\label{sec:stability}
To facilitate the subsequent analysis, we recall the definition of input-to-state stability.
\begin{definition}[Input-to-state stability, {\cite[Definition 4.7]{Khalil.Grizzle2002}}]
	Let $t_0\in\mathbb{R}_{\geq 0}$ denote the initial time. The switched system in~\eqref{eq:ensemble-estimation-error-dynamics} is \emph{input-to-state stable} (ISS) if there exists a class $\mathcal{KL}$ function $\varsigma:\mathbb{R}_{\geq 0}\times\mathbb{R}_{\geq 0}\to\mathbb{R}_{\geq 0}$ and a class $\mathcal{K}$ function $\varphi: \mathbb{R}_{\geq 0}\to\mathbb{R}_{\geq 0}$ such that for any initial state $\tilde{x}(t_0)$ and any bounded input $( \delta_q^\top\delta_q )^{-1}\delta_q^\top\delta_\tau \dot{x}_{\tau} (t)$, the solution $\tilde{x}(t)$ exists for all $t\geq t_0$ and satisfies 
	\begin{align*}
		\lVert \tilde{x}(t) \rVert&\leq \varsigma\left( \lVert \tilde{x}(t_0) \rVert,t-t_0 \right) \\ 
		&+\varphi\Big( \sup_{t_0\leq \kappa \leq t} \lVert ( \delta_q^\top\delta_q )^{-1}\delta_q^\top\delta_\tau \dot{x}_{\tau} (\kappa) \rVert \Big).
	\end{align*}
\end{definition}

The system in~\eqref{eq:ensemble-estimation-error-dynamics} is exponentially ISS if $\varsigma$ is of the form $\varsigma (a,b)=\varkappa_1 ae^{-\varkappa_2 b}$ for $\varkappa_1,\varkappa_2\in\mathbb{R}_{>0}$. Consider the input-free system $\dot{\tilde{z}}=-k_1\delta_{q_{\sigma (t)}}^\top\delta_{q_{\sigma (t)}}\tilde{z}$ and candidate Lyapunov function $V_1(\tilde{z})=\tilde{z}^\top\tilde{z}$ where $V_1:C^0(G_q;\mathcal{F})\to\mathbb{R}_{\geq 0}$. We now state the stability result for the system in~\eqref{eq:ensemble-estimation-error-dynamics}.
\begin{lemma}[{\cite[Theorem 4.1]{Meng.Chen.ea2018}}]\label{lem:observer-stability}
	Let the switching signal $\sigma (t)$ satisfy~\cite[Assumption 2.1]{Meng.Chen.ea2018} and the system in~\eqref{eq:ensemble-estimation-error-dynamics} satisfy~\cite[Assumption 4.1]{Meng.Chen.ea2018}. Then, the switched system in~\eqref{eq:ensemble-estimation-error-dynamics} is exponentially ISS, the function $\varphi$ is of the form
	\begin{align}\label{eq:iss-varphi}
		\varphi(\bar{\dot{x}}_{\tau}) = \lVert \left( \delta_q^\top\delta_q \right)^{-1}\delta_q^\top\delta_\tau  \rVert\frac{2-\sqrt\vartheta}{1-\sqrt\vartheta} B_\sigma \bar{\dot{x}}_{\tau},
	\end{align}
	and there exists $\vartheta\in (0,1)$ which satisfies $V_1(\tilde{z}(t_{k+1}))\leq\vartheta V_1(\tilde{z}(t_k))$ for all $k\in\mathbb{Z}_{\geq 0}$.
\end{lemma}
\begin{proof}
    Assumption~\ref{ass:infinite-sequence} is equivalent to~\cite[Assumption 2.1]{Meng.Chen.ea2018}. Next, we show that~\cite[Assumption 4.1]{Meng.Chen.ea2018} is satisfied. Consider any finite collection of sub-cellular sheaves $\{ \mathcal{F}_1,\ldots,\mathcal{F}_S \}$ whose union cellular sheaf $\mathcal{F}_{\Sigma_S}\triangleq \cup_{s\in\{ 1,\ldots, S \}}\mathcal{F}_s$ satisfies $\delta_{q_{\Sigma_S}}^\top\delta_{q_{\Sigma_S}}\succ 0$. For any finite collection of matrices $\{A_1,\ldots,A_S\}$ satisfying $A_s:\mathbb{R}^\ell\to\mathbb{R}^{m}$, where $\ell\in\mathbb{Z}_{>0}$ and $m\in\mathbb{Z}_{>0}$ for all $s\in\{1,\ldots, S\}$, a vector $\mu\in\mathbb{R}^\ell$ satisfies $\mu\in\mathrm{ker}\ A_1 \cap \ldots \cap \mathrm{ker}\ A_S$ if and only if $\mu\in\mathrm{ker}\ [A_1^\top,\ldots,A_S^\top]^\top$. Then, for any $\mu\in\mathrm{ker}\ A_1 \cap \ldots \cap \mathrm{ker}\ A_S$,
    \begin{align*}
        \sum_{s\in\{1,\ldots,S\}}A_s\mu=0.
    \end{align*}
    By this fact and the relation $\delta_{q_s}^\top\delta_{q_s}=\delta_q^\top \Psi_s\delta_q$ for all $s\in\{1,\ldots, S\}$, $\mu\in\mathrm{ker}\ \delta_{q_1}^\top\delta_{q_1}\cap\ldots\cap\mathrm{ker}\ \delta_{q_S}^\top\delta_{q_S}$ implies $\delta_q^\top \left( \Psi_1 + \ldots +\Psi_S \right)\delta_q \mu =0$. 
    
    Since $\delta_{q_s}^\top\delta_{q_s}\succeq 0$ for all $s\in\{1,\ldots,S\}$,
    \begin{align*}
        \bigcap_{s=1}^S\mathrm{ker}\ \delta_{q_s}^\top\delta_{q_s}=\ker\ \sum_{s=1}^S\delta_{q_s}^\top\delta_{q_s}.
    \end{align*}
    Further, since every edge in the union cellular sheaf occurs in at least one sub-cellular sheaf, we have that for any fixed ordering of $E(G)$, 
    \begin{align*}
        \sum_{s\in\{1,\ldots,S\}}\Psi_s = \mathrm{blkdiag}\Big\{ \Big( c_e \mathbf{1}_{E(G_{\Sigma_S})}(e) I_{\mathcal{F}(e)} \Big)_{e\in E(G)} \Big\},
    \end{align*}
    where $G_{\Sigma_S}\triangleq\cup_{s\in\{1,\ldots, S\}}G_s$ and $c_e\in\mathbb{Z}_{>0}$ is a constant indicating the number of occurrences of a given edge in the sequence. The constant $c_e$ satisfies $1\leq c_e\leq S$ for each $e\in E(G_{\Sigma_S})$. For any $\xi\in C^1(G;\mathcal{F})$, 
    \begin{align*}
        \xi^\top \Psi_{\Sigma_S}\xi = \sum_{e\in E(G_{\Sigma_S})} (\xi)_e^\top (\xi)_e,
    \end{align*}
    and
    \begin{align*}
        \xi^\top \left( \Psi_1 + \ldots + \Psi_S \right)\xi = \sum_{e\in E(G_{\Sigma_S})} c_e (\xi)_e^\top (\xi)_e.
    \end{align*}
    Since $1\leq c_e$ for each $e\in E(G_{\Sigma_S})$,
    \begin{align*}
        \xi^\top \Psi_{\Sigma_S}\xi\leq \xi^\top (\Psi_1 + \ldots + \Psi_S)\xi
    \end{align*}
    holds for all $\xi\in C^1(G;\mathcal{F})$ and $\Psi_{\Sigma_S}\preceq (\Psi_1 + \ldots + \Psi_S)$. The facts that $\Psi_{\Sigma_S}\preceq (\Psi_1 + \ldots + \Psi_S)$ and $\delta_{q_{\Sigma_S}}^\top\delta_{q_{\Sigma_S}}=\delta_q^\top \Psi_{\Sigma_S}\delta_q$ imply 
    \begin{align*}
        \delta_{q_{\Sigma_S}}^\top\delta_{q_{\Sigma_S}} \preceq\sum_{s=1}^S \delta_{q_s}^\top\delta_{q_s}.
    \end{align*}
    Then, since $\delta_{q_{\Sigma_S}}^\top\delta_{q_{\Sigma_S}}\succ 0$, 
    \begin{align*}
        \mathrm{ker}\ \sum_{s=1}^S\delta_{q_s}^\top\delta_{q_s}=\bigcap_{s=1}^S \mathrm{ker}\ \delta_{q_s}^\top\delta_{q_s}=\{ 0 \}.
    \end{align*}
    Thus, by Assumption~\ref{ass:jointly-extensible},
    \begin{align*}
        \bigcap_{\ell =0}^{s_k-1}\mathrm{ker}\ \delta_{q_{\sigma(t_k^\ell)}}^\top \delta_{q_{\sigma(t_k^\ell)}} = \{ 0 \}
    \end{align*}
    for all $k\in\mathbb{Z}_{\geq 0}$ and~\cite[Assumption 4.1]{Meng.Chen.ea2018} is satisfied. Since~\cite[Assumption 2.1]{Meng.Chen.ea2018} and~\cite[Assumption 4.1]{Meng.Chen.ea2018} are satisfied, \cite[Lemma 4.1]{Meng.Chen.ea2018} holds and there exists $\vartheta\in (0,1)$ such that $V_1 (t_{k+1})\leq\vartheta V_1(t_{k})$ for all $k\in\mathbb{Z}_{\geq 0}$. Thus, by~\cite[Theorem 4.1]{Meng.Chen.ea2018} and Assumption~\ref{ass:target-bounds},
    \begin{align}\label{eq:observer-convergence}
        \begin{split}
            \lVert \tilde{x}(t) \rVert &\leq \varkappa_1\lVert \tilde{x}(t_0) \rVert e^{-\varkappa_2 (t-t_0)} + \varphi(\bar{\dot{x}}_\tau),
        \end{split}
    \end{align}
    where $\varphi(\bar{\dot{x}}_\tau)$ is given in~\eqref{eq:iss-varphi} and $\varkappa_1,\varkappa_2\in\mathbb{R}_{>0}$. 
\end{proof}

Consider the candidate Lyapunov function $V_2(\hat{e})=\frac{1}{2}\hat{e}^\top\hat{e}$, where $V_2:C^0 (G_q;\mathcal{F})\to\mathbb{R}_{\geq 0}$. Define the strictly increasing function $\rho:\mathbb{R}_{\geq 0}\to\mathbb{R}_{\geq 0}$ and the class $\mathcal{K}_\infty$ function $\bar{\rho}:\mathbb{R}_{\geq 0}\to \mathbb{R}_{\geq 0}$ satisfying $\bar\rho (a)\triangleq\rho(a) - \rho(0)$ for all $a\in\mathbb{R}_{\geq 0}$. By Assumption~\ref{ass:drift} and Lemma~\ref{lem:observer-stability}, there exists $\bar{f}_{\hat{x}}\in\mathbb{R}_{>0}$ satisfying $\lVert f(\hat{x}(t),t) \rVert\leq \bar{f}_{\hat{x}}$ for all $t\in\mathbb{R}_{\geq t_0}$. To facilitate the subsequent stability analysis, define $\upsilon\in\mathbb{R}_{>0}$ as $\upsilon\triangleq\frac{(\bar{\omega}\sqrt{N}+ \bar{f}_{\hat{x}})^2}{2k_2}$, define the user-selected desired rate of convergence $\lambda\in\mathbb{R}_{>0}$, and define the constant $c_\rho\triangleq \bar{\rho}^{-1}\big(\frac{k_2}{2}-\lambda-\rho (0)\big)\in\mathbb{R}_{>0}$. Next, define the domain 
\begin{align*}
    \mathcal{D}\triangleq\big\{ \mu\in C^0(G_q;\mathcal{F}) \mid \lVert \mu \rVert\leq c_\rho\big\}, 
\end{align*}
the set of initial conditions
\begin{align*}
    \mathcal{S}\triangleq\left\{ \mu\in C^0(G_q;\mathcal{F})\ \Big\vert\  \lVert \mu \rVert < \sqrt{c_\rho^2 -\frac{\upsilon}{\lambda}}  \right\},
\end{align*}
and the ultimate bound
\begin{align*}
    \mathcal{U}\triangleq\left\{ \mu\in C^0(G_q;\mathcal{F})\ \Big\vert\ \lVert \mu \rVert\leq \sqrt{\frac{\upsilon}{\lambda}} \right\}.
\end{align*}
Provided that the sufficient gain condition $\frac{k_2}{2}>\lambda + \rho(\sqrt{\frac{2\upsilon}{\lambda}})$ holds, applying the fact that $\frac{\upsilon}{\lambda}>0$ yields $\mathcal{S}\neq\emptyset$ and $\mathcal{U}\subset\mathcal{S}\subset\mathcal{D}$. By Assumptions~\ref{ass:drift} and~\ref{ass:disturbance}, the vector field on the right hand side of~\eqref{eq:ensemble-regulation-error-dynamics} satisfies the conditions of~\cite[Theorem 3.1]{Khalil.Grizzle2002} and there exists a unique local solution $\hat{e}(t)$. Let $\mathcal{I}\triangleq [t_0,t_{\mathcal{I}})$ denote its maximal interval of existence where $t_{\mathcal{I}}>t_0$. We now state the stability result for the target extension regulation error system. 

\begin{lemma}\label{lem:regulation-stability}
	For target and agent dynamics in~\eqref{eq:target-dynamics} and~\eqref{eq:agent-dynamics}, respectively, the controller in~\eqref{eq:controller} ensures that for any $\hat{e}(t_0)\in\mathcal{S}$, $\hat{e}(t)$ is semi-globally uniformly ultimately bounded and exponentially converges to $\mathcal{U}$, where
    {\small
	\begin{align}\label{eq:regulation-convergence}
		\lVert \hat{e}(t) \rVert\leq\sqrt{ e^{-2\lambda (t-t_0)}\lVert\hat{e}(t_0)\rVert^2+\frac{\upsilon}{\lambda}\left(1-e^{-2\lambda (t-t_0)}\right)},
	\end{align} 
	}for all $t\in\mathbb{R}_{\geq t_0}$ provided that the sufficient gain condition $\frac{k_2}{2}>\lambda+\rho (\sqrt{\frac{2\upsilon}{\lambda}})$ is satisfied. 
\end{lemma}
\begin{proof}
	Using~\eqref{eq:ensemble-regulation-error-dynamics}, taking the derivative of $V_2$ along trajectories of $\dot{\hat{e}}$ yields
	\begin{align}\label{eq:lie-deriv-init}
		\begin{split}
			\dot{V}_2 (\hat{e},t)&= -k_2\hat{e}^\top\hat{e} - \hat{e}^\top\left( f(x,t) - f(\hat{x},t) \right)  \\
			&-\hat{e}^\top \omega (t) - \hat{e}^\top f(\hat{x},t).
		\end{split}
	\end{align}
	Applying Lemma~\ref{lem:observer-stability}, Assumptions~\ref{ass:drift} and~\ref{ass:disturbance}, and the Cauchy-Schwarz inequality gives $-\hat{e}^\top\omega(t)\leq\bar{\omega}\sqrt{N}\lVert \hat{e} \rVert$ and $-\hat{e}^\top f(\hat{x},t)\leq\bar{f}_{\hat{x}}\lVert \hat{e} \rVert$. Applying Assumption~\ref{ass:drift} and the Fundamental Theorem of Calculus to $\lVert f(x,t)-f(\hat{x},t) \rVert$ yields $\lVert f(x,t)-f(\hat{x},t) \rVert\leq \rho \left(\lVert \hat{e} \rVert\right)\lVert\hat{e} \rVert$. Completing the square yields $-\frac{k_2}{2}\lVert \hat{e}\rVert^2 \leq  \upsilon - \big(\bar{\omega}\sqrt{N} + \bar{f}_{\hat{x}} \big)\lVert\hat{e} \rVert$. Then, applying the aforementioned upper bounds, the definition of $\rho$, and the Cauchy-Schwarz inequality to~\eqref{eq:lie-deriv-init} gives
	\begin{align}\label{eq:lie-bounded}
		\dot{V}_2(\hat{e},t)\leq-\Big( \frac{k_2}{2} - \rho\left(\lVert \hat{e} \rVert\right) \Big) \lVert \hat{e} \rVert^2 + \upsilon,
	\end{align}
    for all $t\in\mathcal{I}$ such that $\hat{e}(t)\in\mathcal{D}$.
	
	Next, we show that $\hat{e}(t)$ can be extended to $[t_0,\infty)$. Since $\hat{e}(t)$ is continuous and $\hat{e}(t_0)\in\mathcal{S}$, there exists $t_E >t_0$ such that $\hat{e}(t)\in\mathcal{D}$ for all $t\in[t_0,\min\{t_E,t_{\mathcal{I}} \})$. That is,
    \begin{align*}
        t_E\triangleq \inf\{ t\in [t_0,t_{\mathcal{I}})\ \vert\  \hat{e}(t)\notin \mathcal{D} \}.
    \end{align*}
    Let $\inf\emptyset=\infty$ and assume for contradiction that $t_E < \infty$. Then, $t_E < t_{\mathcal{I}}$ and $\hat{e}(t)\in\mathcal{D}$ for all $t\in [t_0,t_E)$. Using the definitions of $\mathcal{D}$ and $V_2$ to upper bound~\eqref{eq:lie-bounded} and solving the resulting differential inequality gives~\eqref{eq:regulation-convergence} for all $t\in[t_0,t_E)$. Applying the fact that $\hat{e}(t_0)\in\mathcal{S}$ yields $\lVert \hat{e}(t) \rVert < c_\rho$ for all $t\in[t_0,t_E)$. By the continuity of $\hat{e}(t)$, taking the limit of~the aforementioned inequality as $t\to t_E$ yields $\lVert \hat{e}(t_E) \rVert<c_\rho$ and $\hat{e} (t_E)\in\mathrm{int}\ \mathcal{D}$, violating the assumption made for contradiction. Therefore, $t_E = \infty$ and $\hat{e}(t)\in\mathrm{int}\ \mathcal{D}$ for all $t\in\mathcal{I}$. Since $\hat{e}(t)$ remains in the compact set $\mathcal{D}$ for all $t\in\mathcal{I}$, \cite[Theorem 3.3]{Khalil.Grizzle2002} yields that a unique solution $\hat{e}(t)$ exists for all $t\geq t_{0}$ and $\mathcal{I}=[t_{0},\infty)$. Thus, \eqref{eq:regulation-convergence} holds for all $t\in[t_0,\infty)$. 

	Taking the limit superior of~\eqref{eq:regulation-convergence} as $t\to\infty$ yields that $\hat{e}(t)$ converges to $\mathcal{U}$. The rate of convergence is independent of the initial condition $\hat{e}(t_0)$, and thus the convergence is uniform. Since $\bar{\rho}\in\mathcal{K}_\infty,\bar{\rho}^{-1}\in\mathcal{K}_\infty$, then $k_2\to\infty$ implies $c_\rho\to\infty$. Thus, $\mathcal{S},\mathcal{D}$ can be made arbitrarily large through selection of $k_2$ and the stability result is semi-global. By~\eqref{eq:regulation-convergence}, $\hat{e}\in\mathcal{L}_\infty$. By Lemma~\ref{lem:observer-stability}, $\dot{\hat{x}}\in\mathcal{L}_\infty$. Therefore, by Assumption~\ref{ass:control-eff-inv}, $u\in\mathcal{L}_\infty$, where $u\triangleq\oplus_{i\in V(G_q)}u_i$.
\end{proof}
\begin{figure*}[!t]
    \centering
    \includegraphics[width=0.75\linewidth]{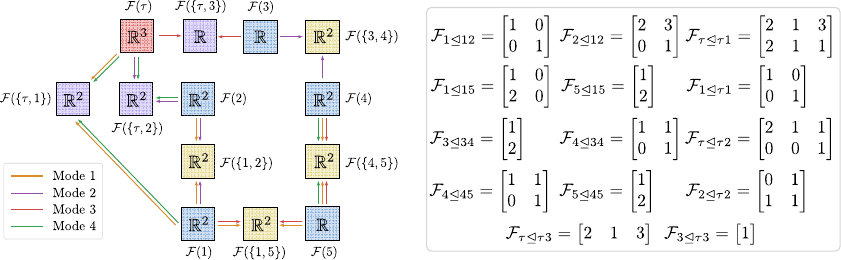}
    \caption{Left: Node stalks, edge stalks,  communication topology, and switching modes for the MAS defined on the cellular sheaf $\mathcal{F}$. Blue squares denote node stalks of agents, yellow squares denote agent-agent edge stalks, the red square denotes the target node stalk, and purple squares denote agent-target edge stalks. Right: Restriction maps of $\mathcal{F}$.}
    \label{fig:simulation-restriction-maps}
\end{figure*}
\noindent We now provide the main result.
\begin{theorem}\label{thm:tracking-error}
	Let Assumptions~\ref{ass:infinite-sequence}-\ref{ass:control-eff-inv} hold and let the conditions of Lemmas~\ref{lem:observer-stability} and~\ref{lem:regulation-stability} be satisfied. Then,
	\begin{align*}\label{eq:tracking-error-bound}
		\limsup_{t\to\infty}\ \lVert e(t) \rVert \leq \lVert \left( \delta_q^\top\delta_q \right)^{-1}\delta_q^\top\delta_\tau \rVert\frac{2-\sqrt{\vartheta}}{1-\sqrt{\vartheta}} B_\sigma \bar{\dot{x}}_{\tau} +\sqrt{\frac{\upsilon}{\lambda}}.
	\end{align*}
\end{theorem}
\begin{proof}
	By the definition of $e$, $\lVert e \rVert=\lVert \tilde{x} + \hat{e} \rVert$. Applying the triangle inequality, the bounds from~\eqref{eq:observer-convergence} and~\eqref{eq:regulation-convergence}, and taking the limit superior as $t\to\infty$ gives the stated result in Theorem~\ref{thm:tracking-error}.
\end{proof}

\section{Simulated Results}\label{sec:simulations}
\begin{figure}[!t]
    \centering
    \includegraphics[width=\linewidth]{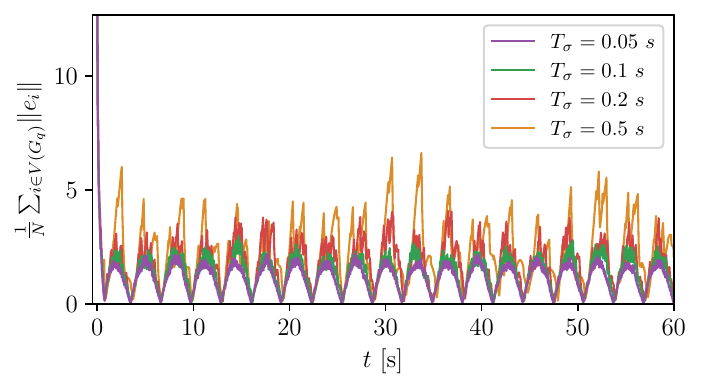}
    \caption{Mean of the normed target tracking error across $4$ trials in which the upper bound of the switching mode interval $(t_k^j-t_k^i)\sim U[0.01,T_\sigma]$ was varied. As $T_\sigma$ increases, more time is spent in disconnected modes, leading to an increase in the mean tracking error norm.}
    \label{fig:mean-tracking-error}
\end{figure}
We validate the temporal cellular sheaf framework empirically for a cellular sheaf MAS having $5$ agents and a switching communication topology. The simulation was conducted for a duration of $t=60\ \mathrm{s}$ with a time step of $\Delta t=0.001\mathrm{s}$. The switching modes, node stalks, edge stalks, and restriction maps of $\mathcal{F}$ are shown in Fig. \ref{fig:simulation-restriction-maps}. The target has dynamics
\begin{align*}
    \dot{x}_\tau=\begin{bmatrix}
        0 & -0.025 & 0.125 \\
        0.025 & 0 & 0.05 \\
        -0.125 & -0.05 & 0
    \end{bmatrix}x_\tau + \sin (t)\mathds{1}_{n_\tau}.
\end{align*}
Each agent has dynamics
\begin{align*}
    \dot{x}_i&=\tanh(n_i x_i) + 0.1n_i\cos(t)\mathds{1}_{n_i} \\
    &+\mathrm{diag}\{1+\cos(x_i)\sin(0.05x_i)+0.1\sin(2t)\mathds{1}_{n_i}\}u_i.
\end{align*}
Each agent's initial position was drawn from the uniform distribution $x_i\sim U[-5,5]^{n_i}$ and the target's initial position was drawn from the uniform distribution $x_\tau\sim U[-4,4]^{n_\tau}$. These initial conditions were fixed across all trials. The nominal cellular sheaf satisfies $\mathrm{Im}\ \delta_\tau\subseteq\mathrm{Im}\ \delta_q$ and $\delta_q^\top\delta_q\succ 0$. The observer gain was selected as $k_1=55$ and the controller gain was selected as $k_2=35$. Across all trials, the switching mode occupancy lower bound was fixed as $\beta_\sigma = 0.01\ \mathrm{s}$. The time between each switching mode was drawn from the uniform distribution $(t_{k}^j-t_k^i)\sim U[0.01,T_\sigma]$ and $T_\sigma\in\mathbb{R}_{>0}$ was varied across trials. Fig.~\ref{fig:mean-tracking-error} demonstrates that increasing $T_\sigma$ leads to degraded tracking error performance. 

\section{Conclusion}
In this work, we introduced a temporal cellular sheaf framework for heterogeneous MASs with switching communication topologies. By organizing these systems into the category $\CellSh$ and using pushout-based gluing constructions, we modeled the evolution of communication and coordination structure across time intervals. Then, using a Lyapunov-based stability analysis, we derived sufficient conditions under which the resulting switching cellular sheaf system achieves target tracking, with the target tracking error converging to a neighborhood of the origin.

\bibliographystyle{ieeetr}
\bibliography{temporal-sheaves}

@article{Bumpus.Leal.ea2025,
  author = {Benjamin Merlin Bumpus and Wilmer Leal and James Fairbanks and Martti Karvonen and Fr{\'e}d{\'e}ric Simard},
  title = {Towards a Unified Theory of Time-Varying Data},
  journal = {Appl. Categ. Struct.},
  year = {2026},
  volume = {34},
  number = {3},
  pages = {23},
  doi = {10.1007/s10485-026-09860-4},
  url = {https://doi.org/10.1007/s10485-026-09860-4},
  issn = {1572-9095}
}

@article{Ni.Cheng2010,
  title = {Leader-following consensus of multi-agent systems under fixed and switching topologies},
  author = {Ni, Wei and Cheng, Daizhan},
  journal = {Syst. Control Lett.},
  volume = {59},
  number = {3-4},
  pages = {209--217},
  year = {2010},
  publisher = {Elsevier}
}

@article{Hong.Hu.ea2006,
  title = {Tracking control for multi-agent consensus with an active leader and variable topology},
  author = {Hong, Yiguang and Hu, Jiangping and Gao, Linxin},
  journal = {Automatica},
  volume = {42},
  number = {7},
  pages = {1177--1182},
  year = {2006},
  publisher = {Elsevier}
}

@book{Khalil.Grizzle2002,
  title = {Nonlinear systems},
  author = {Khalil, Hassan K and Grizzle, Jessy W},
  volume = {3},
  year = {2002},
  publisher = {Prentice Hall Upper Saddle River, NJ}
}

@article{Yu.Xia2012,
  title = {Adaptive consensus of multi-agents in networks with jointly connected topologies},
  author = {Yu, Hui and Xia, Xiaohua},
  journal = {Automatica},
  volume = {48},
  number = {8},
  pages = {1783--1790},
  year = {2012},
  publisher = {Elsevier}
}

@article{Wen.Duan.ea2013,
  title = {Consensus tracking of multi-agent systems with {L}ipschitz-type node dynamics and switching topologies},
  author = {Wen, Guanghui and Duan, Zhisheng and Chen, Guanrong and Yu, Wenwu},
  journal = {IEEE Trans. Circuits Syst.},
  volume = {61},
  number = {2},
  pages = {499--511},
  year = {2013},
  publisher = {IEEE}
}

@article{Xu.Cao.ea2014,
  title = {Leader-following consensus of non-linear multi-agent systems with jointly connected topology},
  author = {Xu, Wenying and Cao, Jinde and Yu, Wenwu and Lu, Jianquan},
  journal = {IET Control Theory Appl.},
  volume = {8},
  number = {6},
  pages = {432--440},
  year = {2014},
  publisher = {Wiley Online Library}
}

@inproceedings{Hanks.Nino.ea2026,
  title = {Heterogeneous multi-agent multi-target tracking using cellular sheaves},
  author = {Hanks, Tyler and Nino, Cristian F and Barcelo, Joana Bou and Copeland, Austin and Dixon, Warren and Fairbanks, James},
  booktitle = {Eur. Control Conf.},
  pages = {1990--1997},
  year = {2026}
}

@phdthesis{Hansen2020,
  title = {Laplacians of cellular sheaves: Theory and applications},
  author = {Hansen, Jakob},
  year = {2020},
  school = {University of Pennsylvania}
}

@article{Meng.Chen.ea2018,
  title = {Consensus of multiagents in switching networks using input-to-state stability of switched systems},
  author = {Meng, Haofei and Chen, Zhiyong and Middleton, Richard},
  journal = {IEEE Trans. Autom. Control},
  volume = {63},
  number = {11},
  pages = {3964--3971},
  year = {2018},
  publisher = {IEEE}
}

@article{Johnstone1999,
  title = {A note on discrete {C}onduch{\'e} fibrations},
  author = {Johnstone, Peter},
  journal = {Theory Appl. Categ.},
  volume = {5},
  number = {1},
  pages = {1--11},
  year = {1999}
}

@inproceedings{Hanks.Riess.ea2025,
  title = {Distributed multi-agent coordination over cellular sheaves},
  author = {Hanks, Tyler and Riess, Hans and Cohen, Samuel and Gross, Trevor and Hale, Matthew and Fairbanks, James},
  booktitle = {Proc. IEEE Conf. Decis. Control},
  pages = {3057--3064},
  year = {2025}
}

@article{Hansen.Ghrist2021,
  title = {Opinion dynamics on discourse sheaves},
  author = {Hansen, Jakob and Ghrist, Robert},
  journal = {SIAM J. Appl. Math.},
  volume = {81},
  number = {5},
  pages = {2033--2060},
  year = {2021},
  publisher = {SIAM}
}

@article{Bosca.Ghrist2026a,
  title = {Selective Adaptation of Beliefs and Communication on Cellular Sheaves},
  author = {Bosca, Vicente and Ghrist, Robert},
  journal = {arXiv preprint arXiv:2601.22431},
  year = {2026}
}

@article{Bosca.Ghrist2026b,
  title = {Neural Networks as Local-to-Global Computations},
  author = {Bosca, Vicente and Ghrist, Robert},
  journal = {arXiv preprint arXiv:2603.14831},
  year = {2026}
}

@book{Edelsbrunner.Harer2010,
  title = {Computational Topology: An Introduction},
  author = {Edelsbrunner, Herbert and Harer, John},
  year = {2010},
  publisher = {Am. Math. Soc.},
  isbn = {978-0-8218-4925-5},
  doi = {10.1090/mbk/069}
}

@article{Leal.Bumpus.ea2026,
  title = {Time-Varying Data as Sheaves: {A}n Invitation to Narratives},
  author = {Wilmer Leal and Benjamin Merlin Bumpus and Jana K. Nickel and Johan García and James Fairbanks and Warren Dixon},
  year = {2026},
  journal = {arXiv preprint arXiv:2609.09056}
}

@phdthesis{Curry2014,
  title = {Sheaves, cosheaves and applications},
  author = {Curry, Justin Michael},
  year = {2014},
  school = {University of Pennsylvania}
}

@article{Ghrist.Riess2020,
  title = {Cellular sheaves of lattices and the {T}arski {L}aplacian},
  author = {Ghrist, Robert and Riess, Hans},
  journal = {arXiv preprint arXiv:2007.04099},
  year = {2020}
}

@article{Hansen.Ghrist2019,
  title = {Toward a spectral theory of cellular sheaves},
  author = {Hansen, Jakob and Ghrist, Robert},
  journal = {J. Appl. Comput. Topol.},
  volume = {3},
  number = {4},
  pages = {315--358},
  year = {2019},
  publisher = {Springer}
}

@article{Trinh.VanNguyen.ea2018,
  title = {Matrix-weighted consensus and its applications},
  author = {Trinh, Minh Hoang and Van Nguyen, Chuong and Lim, Young-Hun and Ahn, Hyo-Sung},
  journal = {Automatica},
  volume = {89},
  pages = {415--419},
  year = {2018},
  publisher = {Elsevier}
}

@inproceedings{Wang.Pan.ea2024,
  title = {Leader-follower consensus on matrix-weighted switching networks},
  author = {Wang, Kai and Pan, Lulu and Shao, Haibin and Li, Dewei and Li, Yuanlong and He, Shaoying},
  booktitle = {Int. Conf. Control Autom.},
  pages = {798--803},
  year = {2024}
}

@article{Hansen2021,
  title = {Expansion in matrix-weighted graphs},
  author = {Hansen, Jakob},
  journal = {Linear Algebra Appl.},
  volume = {630},
  pages = {252--273},
  year = {2021},
  publisher = {Elsevier}
}

@article{Pan.Shao.ea2022,
  title = {Cluster consensus on matrix-weighted switching networks},
  author = {Pan, Lulu and Shao, Haibin and Mesbahi, Mehran and Li, Dewei and Xi, Yugeng},
  journal = {Automatica},
  volume = {141},
  pages = {110308},
  year = {2022},
  publisher = {Elsevier}
}

@inproceedings{Zhang.Pan.ea2024,
  title = {Matrix-scaled consensus on switching networks},
  author = {Zhang, Xin and Pan, Lulu and Shao, Haibin and Li, Dewei and He, Shaoying and Yu, Wenbin},
  booktitle = {Int. Conf. Control Autom. Robot. Vis.},
  pages = {1124--1129},
  year = {2024}
}

@article{Moustafa.Abdelghany2026,
  title = {Finite-Trace Cellular Sheaves for Validation and Estimation of Hybrid Dynamical Systems},
  author = {Moustafa, Ahmed M and Abdelghany, Muhammad B},
  journal = {IEEE Control Syst. Lett.},
  year = {2026},
  publisher = {IEEE}
}

@inproceedings{Zhao.Hanks.ea2026,
  title = {Asynchronous nonlinear sheaf diffusion for multi-agent coordination},
  author = {Zhao, Yichen and Hanks, Tyler and Riess, Hans and Cohen, Samuel and Hale, Matthew and Fairbanks, James},
  booktitle = {Proc. Am. Control Conf.},
  pages = {2459--2466},
  year = {2026}
}

@incollection{Schubert1972,
  title = {Categories},
  author = {Schubert, Horst},
  booktitle = {Categories},
  pages = {1--5},
  year = {1972},
  publisher = {Springer}
}

@article{Felber.Flores.ea2025,
  title = {A sheaf-theoretic characterization of tasks in distributed systems},
  author = {Felber, Stephan and Flores, Bernardo Hummes and Galeana, Hugo Rincon},
  journal = {arXiv preprint arXiv:2503.02556},
  year = {2025}
}

@article{Zhao.Zelazo2016,
  title = {Localizability and distributed protocols for bearing-based network localization in arbitrary dimensions},
  author = {Zhao, Shiyu and Zelazo, Daniel},
  journal = {Automatica},
  volume = {69},
  pages = {334--341},
  year = {2016}
}

@inproceedings{VanTran.Park.ea2018,
  title = {Bearing-based formation control via distributed position estimation},
  author = {Van Tran, Quoc and Park, Shin-Hyun and Ahn, Hyo-Sung},
  booktitle = {IEEE Conf. Control Technol. Appl.},
  pages = {658--663},
  year = {2018}
}

@article{Cheng.Yan.ea2026,
  title = {Bearing-based formation control of higher-order multi-agent systems over periodic switching graphs},
  author = {Cheng, Haoshu and Yan, Yamin and Liu, Tao},
  journal = {Syst. Control Lett.},
  volume = {208},
  pages = {106332},
  year = {2026}
}

@article{Tang.Loria2022,
  title = {Localization and tracking control of autonomous vehicles in time-varying bearing formation},
  author = {Tang, Zhiqi and Lor{\'\i}a, Antonio},
  journal = {IEEE Control Syst. Lett.},
  volume = {7},
  pages = {1231--1236},
  year = {2022}
}
\end{document}